**Tackling paper mills requires us to prevent future contamination and clean up the past -**

**the case of the journal *Bioengineered***

*Authors*

René Aquarius[1] (ORCID: 0000-0002-0968-6884)

Elisabeth M. Bik[2] (ORCID: 0000-0001-5477-0324)

David Bimler[3]

Morten P. Oksvold[4]

Kevin Patrick[5]

*Affiliations*

[1]Department of Neurosurgery, Radboud university medical center, Nijmegen, The Netherlands.

[2]Harbers Bik, San Francisco, CA, USA.

[3]Independent researcher, Wellington, New Zealand.

[4]Independent researcher, Oslo, Norway.

[5]Independent researcher, Bellevue, WA, USA.

*Correspondence*

René Aquarius - rene.aquarius@radboudumc.nl

*Conflicts of interest statement*

All authors have ongoing collaborations with ImageTwin and have been given free access in return. Both David Bimler and Elisabeth M. Bik receive donations through Patreon to encourage their sleuthing activities. Elisabeth M. Bik receives speaker fees and travel reimbursement to give talks and workshops and does occasional consulting work for research institutions, funders, and publishers.

*Funding statement*

No specific funding was received for this work.

*Author contributions*

RA: conceptualization, data curation, formal analysis, investigation, methodology, supervision, validation, writing – original draft, writing – review and editing.

EMB: conceptualization, data curation, investigation, validation, writing – original draft, writing – review and editing.

DB: conceptualization, data curation, investigation, validation, writing – original draft, writing – review and editing.

MPO: conceptualization, data curation, investigation, validation, writing – original draft, writing – review and editing.

KP: conceptualization, data curation, investigation, validation, writing – original draft, writing – review and editing.

**Key points**

- Taylor & Francis noted that their journal *Bioengineered* was targeted by paper mills.

- All articles published in *Bioengineered* between January 1st 2010 to December 31st 2023 containing the terms "mouse" OR "mice" OR "rat" OR "rats" in title or abstract were assessed for inappropriate image duplication and manipulation using ImageTwin and visual inspection.

- Among the 878 included articles, 226 (25.7%) contained inappropriate image duplication and/or manipulation.

- Actions taken by the publisher appear to have stemmed the tide of new paper mill submissions, but a backlog of contaminated articles remains in the literature.

- Taylor & Francis' lack of editorial action has left the scientific community vulnerable to reading and citing hundreds of problematic articles published in *Bioengineered*.

# Abstract

*Introduction*: Taylor & Francis journal *Bioengineered* has been targeted by paper mills. The goal of this study is to identify problematic articles published in Bioengineered during the period 2010 to 2024.

*Methods*: Dimensions was used to search for articles that contained the terms "mouse" OR "mice" OR "rat" OR "rats" in title or abstract, published in *Bioengineered* between January 1st 2010 to December 31st 2024. All articles were assessed by eye and by using software to detect inappropriate image duplication and manipulation. An article was classified as problematic if it contained inappropriate image duplication or manipulation or had been previously retracted. Problematic articles were reported on PubPeer by the authors, if they had not been reported previously. All included articles were assessed for post-publication editorial decisions.

*Results*: We have excluded all articles published in 2024 from further analysis, as these were all retraction notices. We assessed the remaining 878 articles, of which 226 (25.7%) were identified as problematic, of which 35 had been previously retracted. One retracted article was later de-retracted. One article received a correction. None of the included articles received an expression of concern or the Taylor & Francis "under investigation" pop-up.

*Conclusions*: Taylor & Francis' lack of editorial action has left the scientific community vulnerable to reading and citing hundreds of problematic articles published in Bioengineered. To uphold scientific integrity, Taylor & Francis should use the findings of this study as a starting point to systematically identify all compromised articles in Bioengineered and take appropriate editorial action.

**Introduction**

Paper mills are manipulating the scientific record by selling authorship of poor-quality or fabricated manuscripts (Abalkina et al., 2025). Services provided by paper mills may include the sale of manuscripts, citation schemes, fake peer review, and data fabrication (Christopher, 2021; Else & Van Noorden, 2021; Parker et al., 2024). Because paper mill manuscripts often are based on fabricated data, they can produce high numbers of articles in a short time span. When paper mills target a scientific journal, it can lead to a spectacular increase in the number of published articles in that journal (Byrne & Christopher, 2020). An example of such an increase can be seen in the Taylor & Francis journal *Bioengineered*.

*Bioengineered* saw a ten-fold increase in the number of published articles in 2021 and 2022 compared to the previous years (see Figure 1). Taylor & Francis acknowledged in a [blog post](#) that *Bioengineered* was targeted by paper mills and stopped publishing paper mill produced articles from the start of 2023.

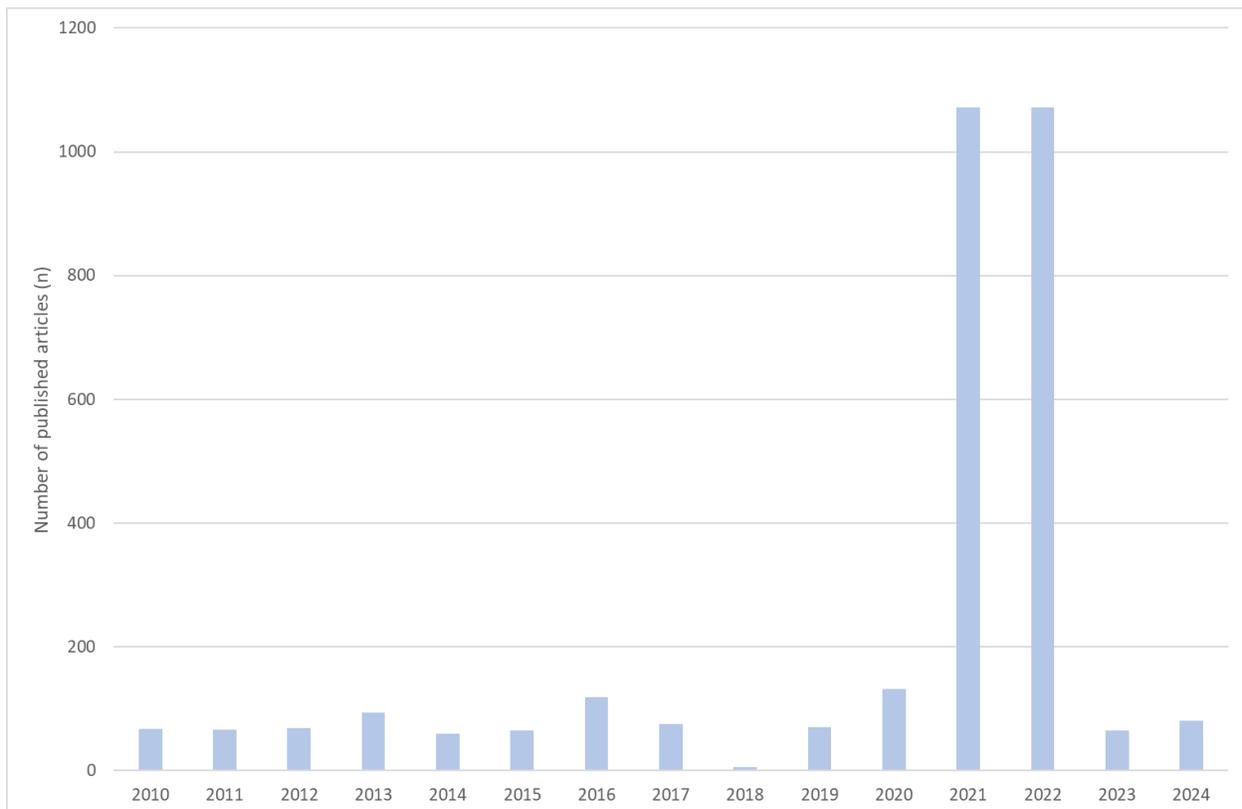

*Figure 1: Number of published articles in Bioengineered from its inception in 2010 until the end of 2024.*
*Source: Dimension.ai.*

Although Taylor & Francis has seemingly been successful in stopping the publication of large amounts of new paper mill articles, they have not retracted all the problematic articles that have been published in the years 2021 and 2022. This is puzzling, as they have clearly stated in their [blog post](#) that they noticed issues in Bioengineered since early 2021, which is now 4 years ago.

The goal of this study was to identify problematic articles published in *Bioengineered* during the period 2010-2024. In our search, we focused on articles with photographic images, because duplications in photos within or across papers are objective indicators of sloppiness or an intention to mislead. Specifically, duplicated or overlapping photos found in multiple papers from different groups of authors at different institutions point towards a common source selling data or complete papers, a feature associated with paper mill activity (Parker et al., 2024). We have also assess how often an editorial decision has been made to correct or retract problematic articles that were published during these years (2010-2024).

**Methods**

*Search*

We performed several searches in Dimensions (Dimensions.ai, Digital Science, London, UK) on December 16 2024, March 19 2025 and March 20 2025. Our searches included articles published in the journal *Bioengineered* between January 1st 2010 to December 31st 2024, that contained the terms "mouse" OR "mice" OR "rat" OR "rats" in title or abstract. We exported the results of our searches as Excel files (Microsoft, Redmond, USA) via the Dimensions.ai interface, which we later merged into one master Excel file. The decision to focus on articles containing these terms was made because they frequently include figures such as histology images and Western immunoblots, which are suitable for assessing inappropriate image duplication or manipulation. Retraction notices were excluded from the analysis.

*Identification of problematic articles*

All authors independently evaluated the included studies using ImageTwin (ImageTwin AI GmbH, Vienna, Austria) and visual inspection between December 16 2024 and March 23 2025. ImageTwin is a software tool designed to detect (partial) overlaps within and between figures in scientific articles. It also identifies (partially) overlapping figures across different articles using its database of over 75 million scientific images (Oza, 2023). In order to keep the number of false-positive findings to a minimum, all image-related issues flagged by ImageTwin had to be visually confirmed by at least one of the authors. When in doubt about a finding, it was discussed with at least one other author. When still in doubt after discussion, the issue would not be classified as problematic.

An article was classified as problematic if it contained inappropriate image duplication or manipulation or had been previously retracted.

*Reporting of problematic articles*

Articles with identified image-related issues were reported on PubPeer (Pubpeer.com, The PubPeer Foundation, California, USA) (Barbour & Stell, 2020) unless they had already been flagged by others. Additionally, we documented other types of issues, such as plagiarism, tortured phrases, mismatched primers, incorrect methods on PubPeer when encountered. However, articles with only these non-image-related issues were not classified as problematic for the purposes of this study.

To ensure completeness, one author (RA) re-evaluated all included articles on March 23 2025, to check for any new reports on PubPeer.

*Classification of image problems*

Image-related issues were classified into three categories:

- Image duplication within the same figure (either within or between panels of the same figure).

- Image duplication between figures within the same article.

- Image duplication between different articles.

*Editorial decisions*

One author (RA) accessed all included articles through the Taylor & Francis website on March 23 and March 24 2025 to assess if any article had been retracted or corrected, or if any article received an expression of concern or the Taylor & Francis "under investigation" pop-up (Kincaid, 2024).

*Outcome measures*

We analysed three outcome measures:

1. The number of included articles relative to the total number of articles published in *Bioengineered*, categorized per year.

2. The proportion of included articles identified as containing problematic images, categorized per year.

3. Any editorial decisions made for included articles (eg. correction, expression of concern, or retraction).

**Results**

*Included articles*

Our search identified a total of 885 articles. After removal of 7 retraction notices, all published in 2024, 878 articles remained for further assessment. Because the 7 excluded retraction notices comprised all the articles from 2024 identified by our search, we decided to exclude 2024 from our analysis in its entirety (Table 1, and Supplementary file 1).

Table 1: number of articles assessed and number of problematic articles identified.

| Year | Articles published – n | Articles included in our sample – n (%) | Problematic articles included in our sample – n (%) |
|---|---|---|---|
| 2010 | 67 | 10 (14.9%) | 1 (10.0%) |
| 2011 | 66 | 7 (10.6%) | 0 (0.0%) |
| 2012 | 69 | 8 (8.3%) | 0 (0.0%) |
| 2013 | 94 | 9 (9.6%) | 1 (11.1%) |
| 2014 | 60 | 6 (10.0%) | 0 (0.0%) |
| 2015 | 64 | 3 (4.7%) | 0 (0.0%) |
| 2016 | 119 | 11 (9.2%) | 1 (9.1%) |
| 2017 | 75 | 1 (1.3%) | 0 (0.0%) |
| 2018 | 6 | 0 (0.0%) | N/A |
| 2019 | 70 | 14 (20.0%) | 5 (35.7%) |
| 2020 | 131 | 22 (16.8%) | 11 (50.0%) |
| 2021 | 1072 | 391 (36.5%) | 101 (25.8%) |
| 2022 | 1072 | 394 (36.8%) | 105 (26.6%) |
| 2023 | 64 | 2 (3.1%) | 1 (50.0%) |
| **TOTAL** | **3029** | **878** | **226** |

*Problematic articles*

Among the 878 included articles, 226 (25.7%) were identified as problematic (see Table 1, Figure 2 and Supplementary file 1). Of these, 194 contained image-related issues and 35 had been previously retracted (not all retracted articles contained image-related issues).

The 194 articles with image-related issues were categorized as follows:

- Image duplication within the same figure - 78 articles;
- Image duplication between figures within the same article - 32 articles;
- Image duplication between different articles - 122 articles;

Some articles contain several types of image-related issues, which is indicated in Supplementary file 1.

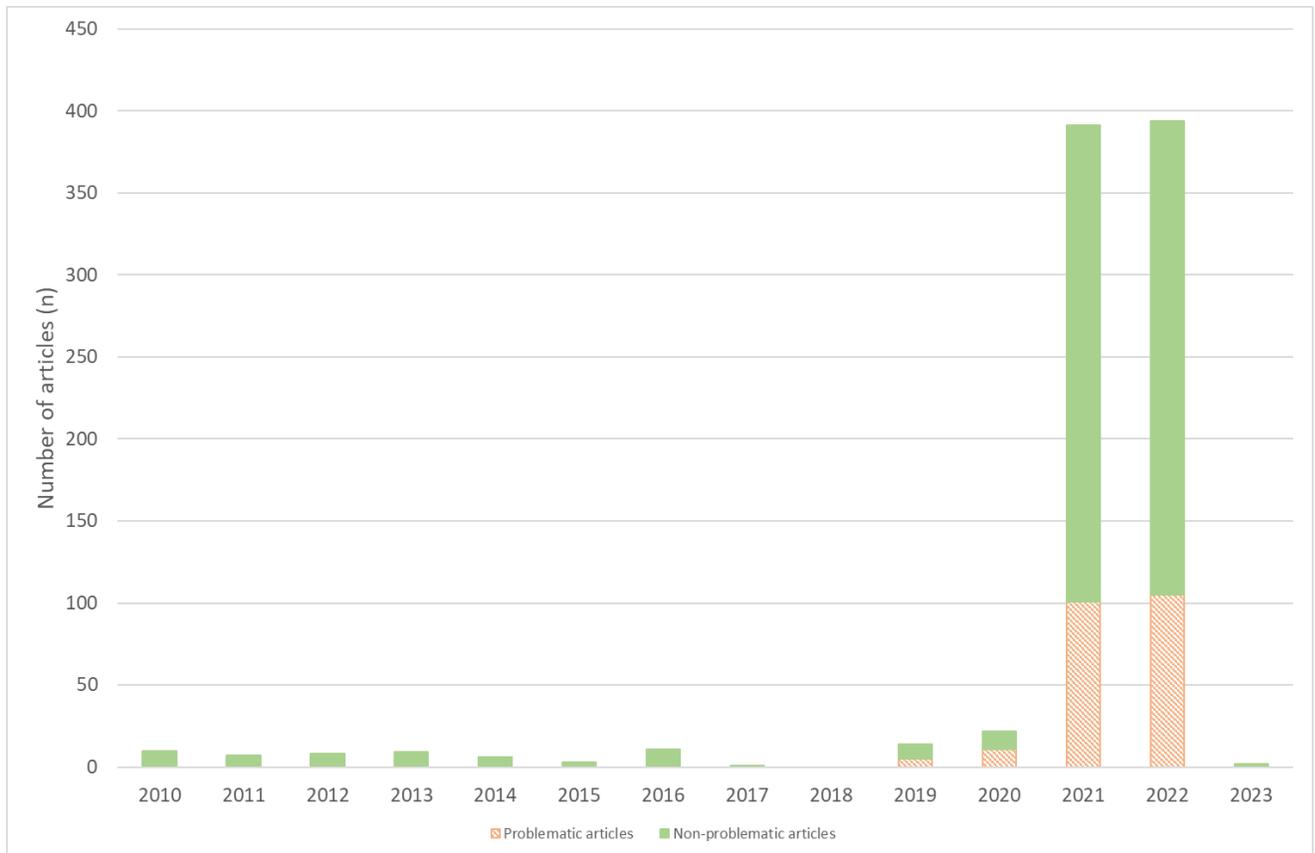

*Figure 2: Included Bioengineered articles in our sample, divided in problematic articles (red, dashed bars) and non-problematic articles (green, solid bars).*

*Number of editorial decisions taken by Bioengineered*

Of all the included articles, a total of 35 articles were retracted. One article (DOI 10.1080/21655979.2021.1987083) had previously been retracted in January 2024, but the retraction was nullified in March 2024 (Supplementary File 1). One article had been previously corrected (DOI 10.1080/21655979.2016.1238534). For both these articles additional issues were found and both should be further investigated by Taylor & Francis.

None of the included articles received an expression of concern or the Taylor & Francis "under investigation" pop-up (Kincaid, 2024).

**Discussion**

Our study identified  a large number of problematic articles (226 out of 878, 25.7%) in the Taylor & Francis journal *Bioengineered*.

*Shift in publication trends*

It seems as if Bioengineered was already targeted by paper mills before 2021. In the years prior to 2019, the number of articles containing the terms "mouse", "mice", "rat" or "rats" in title or abstract always remained under 15% of the total number of published articles. However, in 2019 and 2020 this increased to over 15% and in 2021 and 2022 the numbers even surpassed 35%. This suggests that preclinical animal studies may have been specifically targeted by paper mills.

*Issues preceding 2021*

Our results suggest that paper mills -or other authors producing sloppy or potentially fraudulent research- had already targeted *Bioengineered* before the surge in accepted articles in 2021 and 2022. The relatively stable number of publications in 2010-2020 suggests that paper mills may have been "testing the waters" to gauge what could pass peer review. Possibly, multiple additional paper mills found their way to *Bioengineered* early 2021, explaining the sudden increase in number of publications. However, other explanations are also possible. Maybe paper mills were waiting for editors to recognize that many authors were eager to publish in *Bioengineered*, leading to editorial policy changes, which allowed for steep increases in the number of published articles. Or maybe paper mills were trying to influence editors with financial rewards to manipulate the manuscript acceptance process, which has been described previously (Joelving, 2024).

*Underestimated scope of the problem*

Our study provides a conservative estimate of the actual number of problematic articles in *Bioengineered*. Beyond the 226 flagged articles, we identified an additional 67 flagged articles with other serious issues, such as tortured phrases, mismatched primers and incorrect methods.

Furthermore, a troubling pattern emerged: many articles appeared questionable even in the absence of inappropriate image duplication or manipulation. For example, many articles contained Western immunoblots that apparently did not contain any apparent duplications, but that looked unrealistic. These immunoblots often contained band shapes resembling those observed in the 'Tadpole paper mill' (Bik, 2020; Byrne & Christopher, 2020), but without the repetitive backgrounds. Such papers were not listed in our analysis as problematic because of lack of duplications or other objective problems. However, the unnatural pattern of bands suggested they could have been digitally generated using algorithmic methods.

*Similarities in geography and title structures*

We were struck by the prevalence of author affiliations with (regional) hospitals and universities in China, raising the question why there was such a lack of geographic diversity. Additionally, article titles and image lay-outs often followed highly similar templates, with titles commonly structured as: [*Non-coding RNA*] + [*simple present tense*] + [*disease*] + [*by/through/via*] + [*present continuous tense*] + [*pathway/process*].

Some examples include:

- LncRNA SNHG1 promotes tumor progression and cisplatin resistance through epigenetically silencing miR-381 in breast cancer
- MiR-211 protects cerebral ischemia/reperfusion injury by inhibiting cell apoptosis
- Downregulated circular RNA hsa_circ_0005797 inhibits endometrial cancer by modulating microRNA-298/Catenin delta 1 signaling
- Circ_0017639 facilitates proliferative, migratory, and invasive potential of non-small cell lung cancer (NSCLC) cells via PI3K/AKT signaling pathway
- Human umbilical cord-mesenchymal stem cells-derived exosomes carrying microRNA-15a-5p possess therapeutic effects on Wilms tumor via regulating septin 2

The absence of geographical diversity, combined with the formulaic writing style, and recurring image layouts, strongly suggests coordinated efforts and template-based writing - likely by paper mills.

*Taylor & Francis has not fully addressed the paper mill problem*

In a [blog post](#) on the Taylor & Francis website, Todd Hummel (Taylor & Francis Global Publishing Director, STM) stated that *Bioengineered* had "*overcome the paper mill problem*". However, their actions make it clear that they are wholly focused on preventing publication of new articles from paper mills, by methods such as integrity checks at submission and vetting of reviewers. While we acknowledge the publisher's effort to improve the submission, peer-review and publication process, our findings indicate that they have not sufficiently addressed problematic articles already published.

One reason may be because it is much harder to retract an already-published paper than to trap paper mill articles at the point of submission. Journal staff need the expertise to recognize when an image problem is highly unlikely to be due to "honest error". They need to communicate with authors regarding retraction, and may encounter authors who are litigious or non-responsive. We understand that large-scale, journal-wide investigations are complex, time-consuming and difficult for all parties involved. However, it remains unclear whether Taylor & Francis is actively investigating these articles. None of the flagged articles display the "under investigation" pop-up, which could serve as a simple yet effective way to alert readers and researchers (Kincaid, 2024). Given that Taylor & Francis generates hundreds of millions of pounds in annual revenue, the publisher has both the resources and the responsibility to systematically investigate *Bioengineered's* archives and retract compromised articles.

There are two reasons why it is important to retract published paper mill articles. First, these articles destroy the integrity of the publication record. They may get cited and find their way into systematic reviews, subverting attempts to integrate the literature. Second, retraction of published articles undermines the business model of paper mills. If it happens often enough, paying customers may be harder to find, because they will hear of others who have paid to have an article published, only to have it later retracted. The customer has no means of getting their money back, and instead of the benefit of a journal publication, they find themselves associated with the stigma of enforced retraction.

*Consequences of Bioengineered's decline*

The issues at *Bioengineered* have had significant repercussions. The journal is now classified as category 0 (not approved) by the Norwegian Register for Scientific Journals. In addition, the journal was listed on the early warning journal list due to papermill activity, by the National Science Library of the Chinese Academy of Sciences in February 2024. Another major concern is that nearly all of the problematic articles continue to be cited. Only 3 of the 226 flagged articles have not been cited. The remaining 223 were cited between 1 and 117 times (Supplementary File 1). Many researchers may be unaware of what has transpired at *Bioengineered*, meaning these problematic articles are still being referenced in scientific literature.

*Limitations*

Our study has several limitations:

1) Sampling bias.

   First, we have only investigated a subset of *Bioengineered's* publications, specifically articles related to rodent studies. Paper mills may have targeted these types of studies more heavily, but other disciplines within the journal may also be affected. Our decision to focus on rodent-related articles means other problematic articles may remain undetected.

2) Potential false positives and false negatives.

   We took great care to minimize false positives by focusing on image manipulations that are relatively easy for readers to verify (see PubPeer links in Supplementary File 1) However, false negatives are possible - some problematic articles may have escaped detection due to more subtle manipulation techniques. Many articles had highly similar figure styles, suggesting common authorship or outsourced manuscript preparation, but not all contained explicit evidence of image duplication.

3) Paper mill attribution.

   In some cases, we have strong evidence of paper mill involvement - for example different authors using the same figure in multiple articles. However, we do not have a clear connection to paper mills for all problematic articles. Thus, it is possible that some articles come from a different type of source.

*The path ahead*

After publishing the manuscript on a pre-print server, we will submit it to *Bioengineered* to give Taylor & Francis a chance to inform their readers about the issues that we have identified in *Bioengineered*. Our work serves as an example of how the work of scientists can be used as input for publishers to improve the scientific record. As previously stated, "*publishers, journals, researchers, and institutions must work together and show courage and determination to recognize and ultimately reject the worthless magic of papers created without experiments*" (Byrne & Christopher, 2020).

**Conclusions**

Taylor & Francis' lack of editorial action has left the scientific community vulnerable to reading and citing hundreds of problematic articles published in *Bioengineered*. To uphold scientific integrity, Taylor & Francis should use the findings of this study as a starting point to systematically identify all compromised articles in *Bioengineered* and take appropriate editorial action.

**Acknowledgements**

We thank Dorothy Bishop for reading our manuscript and giving valuable feedback. We thank all (anonymous) science sleuths who have reported issues on *Bioengineered*: your work is much appreciated.

**Supplementary File**

Supplementary data sheet available [here](#).